\documentclass[]{spie}  

\usepackage{amsmath,amsfonts,amssymb}
\usepackage{graphicx}
\usepackage[colorlinks=true, allcolors=blue]{hyperref}
\usepackage{multirow}
\usepackage{astro_bib_macro}

\title{Apodized Pupil Lyot Coronagraphs designs for future segmented space telescopes}

\author{Kathryn St.Laurent\supit{a}, Kevin Fogarty\supit{a}, Neil T. Zimmerman\supit{b}, Mamadou N'Diaye\supit{c}, Christopher C. Stark\supit{a}, Johan Mazoyer\supit{a,d} Anand Sivaramakrishnan\supit{a,d}, Laurent Pueyo\supit{a}, Stuart Shaklan\supit{e}, Robert Vanderbei\supit{f}, R\'{e}mi Soummer\supit{a}
\skiplinehalf
\supit{a} Space Telescope Science Institute, 3700 San Martin Drive, Baltimore, MD 21218, USA\\
\supit{b} NASA Goddard Space Center, 8800 Greenbelt Rd, Greenbelt, MD 20771, USA\\
\supit{c} Observatoire de Nice C\^{o}te d\'{}Azur, Boulevard de l\'{}Observatoire, 06304 Nice, France\\
\supit{d} John Hopkins University, 3400 North Charles Street, Baltimore, MD 21218, USA
\supit{e} Jet Propulsion Laboratory, 4800 Oak Grove Dr, Pasadena, CA 91109, USA\\
\supit{f} Princeton University, 98 Charlton St, Princeton, NJ 08540, USA
}

\authorinfo{Further author information, send correspondence to Kathryn St.Laurent: E-mail: kstlaurent@stsci.edu, Telephone: 1 410 338 6764}

\begin{document} 
\maketitle

\begin{abstract}
A coronagraphic starlight suppression system situated on a future flagship space observatory offers a promising avenue to image Earth-like exoplanets and search for biomarkers in their atmospheric spectra. One NASA mission concept that could serve as the platform to realize this scientific breakthrough is the Large UV/Optical/IR Surveyor (LUVOIR). Such a mission would also address a broad range of topics in astrophysics with a multi-wavelength suite of instruments. The apodized pupil Lyot coronagraph (APLC) is one of several coronagraph design families that the community is assessing as part of NASA’s Exoplanet Exploration Program Segmented aperture coronagraph design and analysis (SCDA) team. The APLC is a Lyot-style coronagraph that suppresses starlight through a series of amplitude operations on the on-axis field. Given a suite of seven plausible segmented telescope apertures, we have developed an object-oriented software toolkit to automate the exploration of thousands of APLC design parameter combinations. This has enabled us to empirically establish relationships between planet throughput and telescope aperture geometry, inner working angle, bandwidth, and contrast level. In parallel with the parameter space exploration, we have investigated several strategies to improve the robustness of APLC designs to fabrication and alignment errors. We also investigate the combination of APLC with wavefront control or complex focal plane masks to improve inner working angle and throughput. Preliminary scientific yield evaluations based on design reference mission simulations indicate the APLC is a very competitive concept for surveying the local exoEarth population with a mission like LUVOIR.
\end{abstract}


\section{INTRODUCTION}
\label{sec:INTRODUCTION}

The Exoplanet Exploration Program (ExEP) has organized a technical study, Segmented Coronagraph Design and Analysis (SCDA) \cite{siegler_shaklan_2016} \cite{scda_aps} . This study seeks to understand the working capability of various coronagraph designs with segmented and obscured telescope apertures, in support of possible future mission concepts being studied by NASA in preparation for the 2020 Decadal Survey. The overall goal is to image terrestrial analogs in the habitable zone of nearby stars. The results of the SCDA effort has directly informed the mission concept study being carried out by the LUVOIR Science and Technology Definition Team.

The apodized pupil Lyot coronagraph (APLC) \cite{2002A&A...389..334A} \cite{2003EAS.....8...93S} \cite{2005ApJ...618L.161S} \cite{2009ApJ...695..695S} \cite{2011ApJ...729..144S} is one of several coronagraph design families that SCDA has assessed, in particular the recent hybrids replacing graded-transmission apodizers with binary-transmission shaped pupils. For further details, see Zimmerman 2016\cite{2016SPIE.9904E..1YZ}.

\section{APLC optimization strategy}
\label{param_space}

APLCs are sensitive to several parameters: telescope aperture shape and segmentation, central obscuration, lyot stop shape and size, focal place mask size, dark hole size, and bandwidth. We utilize both large ($>$100 designs) and small scale design surveys to study this multi-dimensional parameter space. Our large surveys explore many possible telescope geometries and give us an understanding of general properties and scaling of aperture features, such as central obscuration. Our small surveys focus on one or two telescope apertures. They are structured to answer specific questions -- how do secondary mirror support struts impact throughput, can we make a design robust to lyot stop misalignment, etc.

We have greatly expanded our explored parameter space since Zimmerman et al. 2016\cite{2016SPIE.9904E..1YZ}. In particular we have increased our coverage of inner working angle and telescope aperture geometry. Tens of thousands of designs have been produced and analyzed, exploring telescope aperture geometry, central obscuration, focal plane mask size, dark zone size, dark zone bandpass, and Lyot stop geometry. Our ultimate measure of a given design's success is the number of exoEarths it can theoretically detect, discussed in Section \ref{sci_yields}.

\subsection{Linear optimization and software development}

Our apodizer optimization method builds on methods originally devised for the shaped pupil \cite{2003ApJ...599..686V} \cite{2003SPIE.4860..240K} \cite{2011OExpr..1926796C}. A linear program maximises transmission of the apodizer/shaped pupil given a contrast goal and bounds on the dark zone. This linear optimization approach was recently adapted to the APLC design case by expanding the numerical propagation model to include the masks at intermediate planes \cite{0004-637X-818-2-163}. A detailed algebraic description of the linear Lyot coronagraph propagation model is given in the appendix of Zimmerman et al. (2016)\cite{2016JATIS...2a1012Z}, presented for several cases of pupil symmetry and focal plane mask geometry. Our linear solver (in AMPL/Gurobi), survey design tools, and analysis tools are bundled together in a python toolkit. The core module, example notebooks, and documentation are publicly hosted at \url{https://github.com/spacetelescope/SCDA}. A full discussion of this toolkit is available in Zimmerman's 2016 SPIE paper \cite{2016SPIE.9904E..1YZ}. 

The wide parameter space open to the SCDA study motivated us to examine the application of a supercomputer to scale up the feasible volume of optimization trials. Among the various computing facilities managed by NASA, we found that the NASA Center for Climate Simulation (NCCS) Discover System, situated at Goddard Space Flight Center, offered the most promising match for our needs, and obtained time allocation on the cluster. Using the Discover server gave us an order of magnitude effective survey speedup compared to our best case operation with 2–3 local high-performance servers. The main limitation of the Discover System is the execution time ceiling: the queue manager imposes a 24-hour cutoff on each job. The larger coronagraph optimization programs of interest to us–those with high spatial resolution or extra linear constraints–typically require several days to complete. Therefore, we continue to use Linux servers at STScI to optimize select high-resolution design cases as needed.

\subsection{Optimization criteria}
\label{throuput_def}

Our solver optimizes a given APLC design for best possible throughput and a user-input contrast goal. Simply defining PSF throughput relative to the energy incident on the primary mirror becomes problematic when comparing the performance across different telescope architectures. To simplify performance comparisons across different collecting areas, we define throughput relative to the energy that would be collected in an unobscured circular area of the same diameter as the primary mirror. The numerator of our throughput metric is the sum of the PSF in a fixed circular photometric aperture of radius 0.7 $\lambda_{0}/D$. This tends to penalize designs with lower angular resolution, which depends on the apodizer transmission pattern and the Lyot stop outer diameter. Total throughput may be used as a reasonable proxy for peak of planet PSF.

\pagebreak
\section{Surveys}

\subsection{SCDA reference apertures}
\label{SCDA_ref}

\begin{figure}[h]
   \begin{center}
   \begin{tabular}{c}
   \includegraphics[height=3cm]{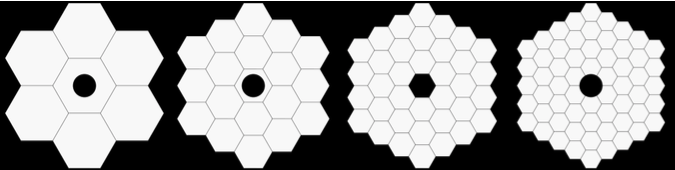} \\
   \includegraphics[height=3cm]{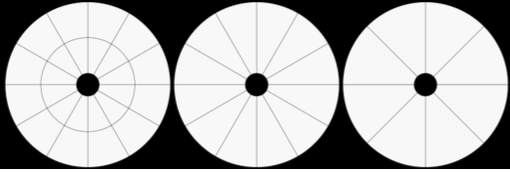}
   \end{tabular}
   \end{center}
   \caption[] 
   { \label{fig:scda_og} 
 SCDA reference apertures \cite{scda_aps}. Apertures include four composed of hexagonal segments, one with keystone segments, and 2 with pie wedges. All are 12 m flat-to-flat or 12 m in diameter with 1.68 m diameter secondary obscurations (except the missing hex segment in the 3-ring hex). All segment edge gaps including edge roll-off are 20 mm wide. Top row: Hex-1, Hex-2, Hex-3, Hex-4. Bottom row: Keystone, Pie-12, Pie-8}
   \end{figure}

We have greatly increased out coverage of inner working angle and telescope aperture geometry since previous efforts \cite{2016SPIE.9904E..1YZ}. We tested approximately 3100 design parameter combinations, varying telescope aperture (see Figure \ref{fig:scda_og}), central obscuration, focal plane mask, dark zone bandpass, and lyot stop size.

Each telescope aperture was matched to a 4-strut secondary support structure chosen to minimize the added obscuration: hexagonal apertures were matched to the X-shaped supports, circular apertures to the cross-shaped supports. For hexagonal APLC designs, the outer diameter (O.D.) of the annular Lyot stop is constrained by the circle inscribed in the aperture perimeter. We fixed the Lyot stop O.D. accordingly for the hexagonal segmentation: 76\%, 82\%, 81\%, and 82\% of pupil diameter for Hex-1, Hex-2, Hex-3, and Hex-4, respectively. For the Keystone/Pie apertures we fixed the Lyot stop O.D. at 90\%. This 90\% O.D. was decided based on a few trial runs that confirmed previous experience that a circular APLC is more efficient with an undersized Lyot stop rather than matching the telescope pupil diameter \cite{2016SPIE.9904E..1YZ}.

The design contrast was fixed at $10^{−10}$ throughout the survey. More specifically, the field in the final image plane was constrained by the optimizer so that the ideal, unaberrated, wavelength-averaged, on-axis intensity pattern is below $10^{−10}$ contrast over the field of view. The outer perimeter of the annular dark zone –  the effective outer working angle – was fixed at 10 \(\lambda/D\) across the survey. The field in the image plane was also constrained in a buffer region interior to the focal plane mask edge, to improve the design robustness to stellar diameter and low-order wavefront aberrations \cite{2015ApJ...799..225N}. For most design parameters we tested two dark zone buffer widths, 0.25 $\lambda/D$ and 0.50 $\lambda/D$ inside of the FPM edge.

Designs with 15\% band-pass require the optimizer to run at 5 wavelengths to constrain the dark zone, versus 4 wavelengths for the 10\% band-pass. The median completion time on NCCS Discover for each 15\% bandwidth apodizer optimization was about 4 hours, but there was a wide scatter in completion times, ranging up to 20 hours.

\begin{figure}
   \begin{center}
   \begin{tabular}{c}
   \includegraphics[height=9.5cm]{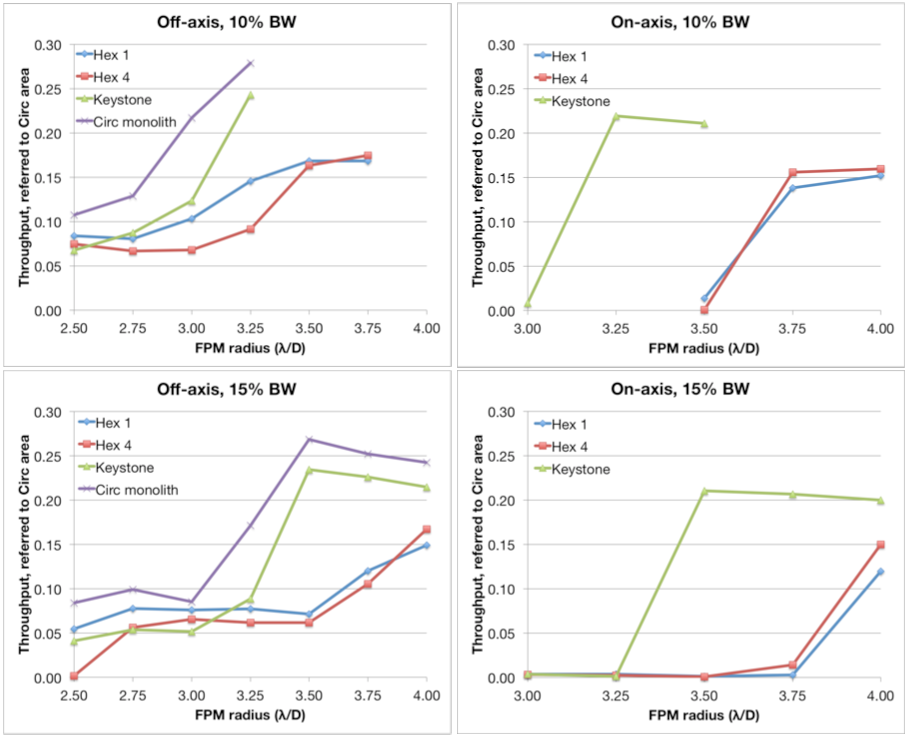}
   \end{tabular}
   \end{center}
   \caption[] 
   { \label{fig:throughput_SCDA} 
PSF throughput of best solutions for each telescope aperture, plotted for each survey bandwidth (10\%, 15\%) and obscuration case (off-axis, on-axis) as a function of focal plane mask radius. To serve as a reference for off-axis performance, the solutions for a clear, monolithic circular aperture are included alongside the SCDA apertures. The results for 2-ring, 3-ring, and 4-ring hexagonal apertures are similar enough that we only represent the 4-ring hexagonal designs; likewise, the Pie wedge-8 and Pie wedge-12 designs are nearly identical in performance to the Keystone designs, so they are not plotted.}
   \end{figure}

In Figure \ref{fig:throughput_SCDA} we plot the throughput of the best designs in our survey for each aperture and bandpass as a function of FPM radius. For the cases of the centrally obscured (on-axis) telescope pupils, shown in the right hand sub-panels, the most striking features are the sharp transitions in throughput as the FPM radius is varied. The performance transition occurs at a significantly smaller radius for the Keystone/Pie apertures than for the Hex apertures. For the 10\% bandpass, the Keystone APLC throughput jumps at FPM radius 3.25 \(\lambda/D\), whereas the throughput of the hexagonal APLC designs remains low until FPM radius 3.75 \(\lambda/D\). The APLC designs for the unobscured (off-axis) apertures, on the other hand, vary more smoothly with FPM size. At FPM radius 3.00 \(\lambda/D\) all of the designs have throughputs $\geq$ 5\%. 

Across all cases of central obscuration and bandwidth, the Keystone/Pie-wedge APLC throughput saturates on a smaller FPM than the hexagonal designs. The advantage is owed to the similarity of these apertures to the ideal, clear circular pupil of the classical APLC, and in particular their circular perimeters. The Keystone APLC mimics the same throughput vs. FPM radius profile of the circular monolith APLC, with a slight throughput attenuation due to apodization of the segment gaps. 

\begin{figure}[h]
   \begin{center}
   \begin{tabular}{c}
   \includegraphics[height=6cm]{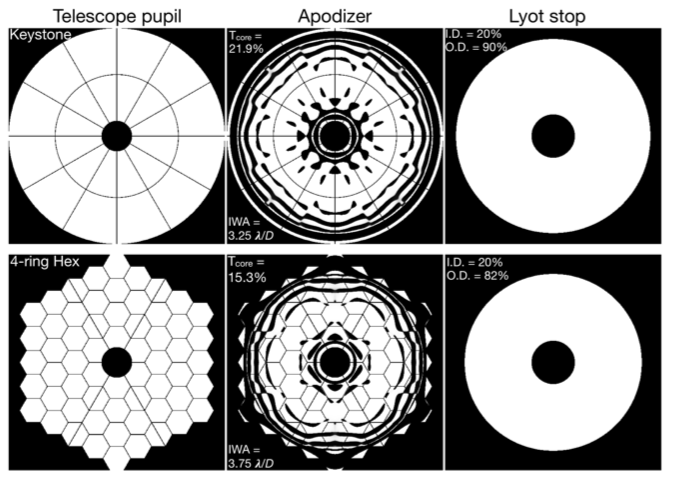}
   \end{tabular}
   \end{center}
   \caption[] 
   { \label{fig:ex_SCDA} 
Two example designs produced by the large design survey, one for an on-axis Keystone primary mirror and one for the on-axis, 4-ring Hexagonal primary. The middle column shows the apodizer solution for each case, and the right hand column shows the corresponding Lyot stop. The $T_{core}$ annotation is the throughput as defined in Section \ref{throuput_def}.}
   \end{figure}

In Figure \ref{fig:ex_SCDA} we display the telescope pupil (downsampled to 500-point array width and rounded to binary), apodizer solution, and Lyot stop of two example APLC designs. These designs correspond to the peak scientific yield metrics described in Section \ref{sci_yields} for the obscured (on-axis) Keystone and 4-ring Hex apertures. The respective FPM radii (and approximate IWA) of these designs are respectively 3.25 $\lambda/D$ and 3.75 $\lambda/D$. Their PSF core throughputs, as defined in Section \ref{throuput_def}, are respectively 21.9\% and 15.3\%.

The throughput advantage of the Keystone APLC is partly owed to the larger primary mirror collecting area. If the telescope aperture diameters are dimensionless and scaled to 1 (flat-to-flat diameter for the Hex-4 case), the collecting areas of the obscured Keystone and Hex-4 primaries are respectively $0.95\pi/4$ and $0.77\pi/4$. But an additional loss factor intrinsic to the coronagraph design arises from the restricted O.D. of the Lyot stop for the hexagonal aperture – as discussed previously, this O.D. is constrained by the inscribed circle in the aperture perimeter. Between the two designs, the ratio of Lyot stop transmission areas that the coronagraph can propagate off-axis light through comes out to about 1.22 in favor of the 90\% O.D. of the Keystone APLC Lyot stop.

\begin{figure}[h]
   \begin{center}
   \begin{tabular}{c}
   \includegraphics[height=4cm]{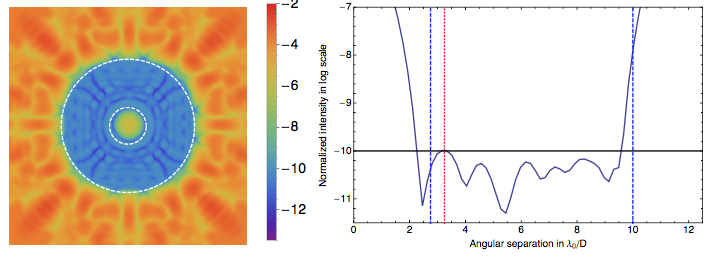}
   \end{tabular}
   \end{center}
   \caption[] 
   { \label{fig:hex4_PSF} 
Band-averaged PSF evaluation for the survey design with peak scientific yield for the obscured Keystone aperture at 10\% bandwidth (same one illustrated in the top row of Figure \ref{fig:ex_SCDA}. The vertical, dashed red line at separation 3.25 $\lambda_{0}/D$ marks the radius of the focal plane mask occulting spot, which determines the inner working angle of the design. The vertical, dashed blue line at separation 2.75 $\lambda_{0}/D$ indicates the radius of the innermost constraint applied during the apodizer optimization to improve the robustness to stellar diameter and jitter.}
   \end{figure}

In Figure \ref{fig:hex4_PSF} we illustrate the on-axis PSF of the same Keystone APLC discussed above. The PSF was computed with an ideal, flat wavefront Fourier propagation, and the intensity pattern was averaged over 11 wavelengths spanning the 10\% design bandpass. The radial PSF profile (averaged over position angle) shows that the design meets the $10^{−10}$ contrast goal, and exceeds it at most angular separations.

For the same Hex-4 APLC design in Figure \ref{fig:ex_SCDA}, in Figure \ref{fig:hex4_PSF} we illustrate how its PSF gradually degrades with increasing angular diameter of the on-axis source. Measured in resolution elements ( $\lambda_{0}/D$ ), the radial profile of the PSF remains within a contrast increment of 1 x $10^{−10}$ of the ideal response of the unresolved point source until the source diameter exceeds 0.2 $\lambda/D$. This roughly corresponds to star of angular diameter 2 mas for a 12-meter aperture at $\lambda_{0} = 600$ nm, or equivalently a solar twin observed at a distance of 5 pc.

\pagebreak   
\subsection{Parameter relationships: Deep dives}
\label{deep_dive}

The design survey described in Section \ref{SCDA_ref} gave us a broad snapshot of segmented aperture APLC performance. We built on that large survey with finer sampled surveys yielding better coverage of FPM radius and Lyot stop dimensions (inner and outer diameter of annulus). This is computationally practical only when we restrict the survey to a small subset of telescope apertures, and permit only 1–2 parameters to vary within a given design batch. This is superseded by our findings described in Section \ref{axi-sym}, but we include the FPM findings as these remain complimentary to our axi-symmetric survey. We also decisively identified the jagged aperture perimeter as the root of the performance disparity between the hexagonal and pie-wedge/keystone apertures.

\subsubsection{FPM radius}
\label{fpm_rad}

\begin{figure}[h]
   \begin{center}
   \begin{tabular}{c}
   \includegraphics[height=7cm]{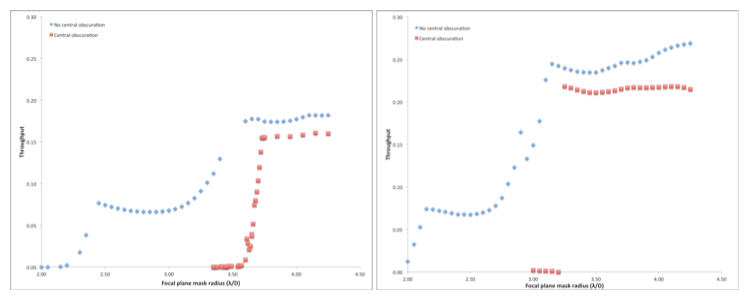}
   \end{tabular}
   \end{center}
   \caption[] 
   { \label{fig:fpm_deep_dive} 
Dependence of APLC throughput on FPM radius for the Hex-4 (left) and Keystone-24 (right) telescope apertures, both with central obscuration (red points) and without central obscuration (blue points).}
   \end{figure}

We wanted to further probe the relationship between FPM radius - a reliable proxy for inner working angle - and throughput. We restricted our study to two representative apertures (Hex-4 and Keystone-24) with and without central obscuration. The results are plotted in Figure \ref{fig:fpm_deep_dive}. In these trials, the bandpass was fixed at 10\% and the Lyot stops were fixed according to the best performing design in each aperture (for the obscured versions these are shown in Figure \ref{fig:ex_SCDA}). As in the large design survey, the contrast goal was fixed at $10^{−10}$. We varied the FPM radius in increments of 0.05 $\lambda_{0}/D$.

For the centrally obscured Hex-4 aperture, we did a second pass to resolve the steep jump in throughput occurring around 3.7 $\lambda_{0}/D$. This time, inside the restricted FPM radius range 3.35–3.75 $\lambda_{0}/D$, we used much finer increments of 0.01 $\lambda_{0}/D$. Finally at this resolution it is possible to trace the transition in IWA between “failed” designs with nearly zero throughput and the high throughput plateau. The finely resolved profile suggests that at this 10\% bandpass, the Hex-4 APLC is optimally paired with a FPM with radius 3.75 $\lambda_{0}/D$. The right hand side of Figure \ref{fig:fpm_deep_dive} shows, as noted before, that for the Keystone-24 aperture this performance transition occurs at a smaller radius of 3.25 $\lambda_{0}/D$.

When the central obscuration is removed, the throughput-IWA curve exhibits a two-tiered drop from the large IWA plateau. The results of our DRM yield trials suggest there is little advantage to operating in the intermediate region of this curve. The drop in throughput, reaching down to ∼ 7\% for FPM radii in the range 2.25–3.25 $\lambda_{0}/D$, is too large to provide a net increase in the detection tally.

\subsubsection{Telescope aperture perimeter}
\label{telap_perim}

During previous discussions, we have hypothesized that the systematic performance disparity between the hexagonal and keystone/pie-wedge apertures is due to the perimeter shape of the primary mirror. Any deviation of the hexagonal aperture perimeter from a circle introduces new diffraction features that the apodizer must compensate for, leading to newly blocked regions in the pupil that would not occur in axisymmetric, circular shaped pupil and APLC solutions \cite{2003ApJ...599..686V} \cite{2015ApJ...799..225N} \cite{2016SPIE.9904E..1YZ}.

\begin{figure}[h]
   \begin{center}
   \begin{tabular}{c}
   \includegraphics[height=5cm]{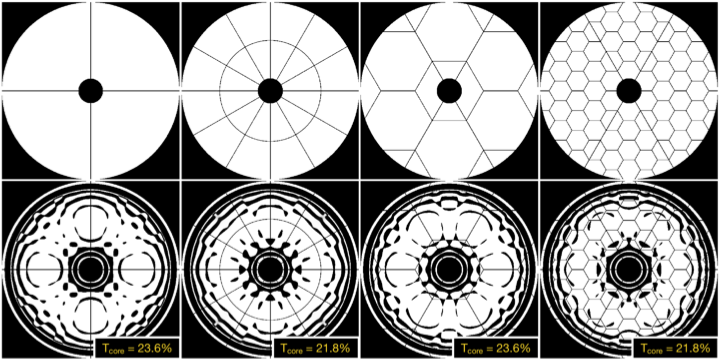}
   \end{tabular}
   \end{center}
   \caption[] 
   { \label{fig:tel_ap_perim} 
Comparison of obscured telescope apertures with circular perimeters and their corresponding apodizer solutions with fixed dark zone parameters (contrast $10^{-10}$ over 10\% bandpass and FPM radius 3.25 $\lambda_{0}/D$, matching the peak DRM yield Keystone design described in Section \ref{SCDA_ref} and Table \ref{table:scda_ref_sy}). From left to right, these are the circular monolith aperture, the keystone aperture, the filled perimeter Hex-1 aperture, and the filled perimeter Hex-4 aperture. The $T_{core}$ value annotated for each apodizer is the PSF core throughput as defined in Section \ref{throuput_def}}
   \end{figure}

We were recently able to verify this hypothesis by re-designing APLCs for modified hexagonal reference apertures, in which the mirror layout pattern continues beyond the original number of whole segment rings, effectively “filling in” the aperture edge to form a circular perimeter. In the top row of Figure \ref{fig:tel_ap_perim} we display the new versions of the centrally obscured Hex-1 and Hex-4 apertures, next to the circular monolith and Keystone-24 apertures, which serve as reference benchmarks from the design survey in Section \ref{SCDA_ref}.

We produced apodizers for these filled-perimeter hexagonal apertures for a $10^{-10}$ contrast, 10\% bandpass, using the same peak-performing FPM and Lyot stop dimensions identified for the obscured Keystone-24 aperture (FPM radius 3.25 $\lambda_{0}/D$; Lyot stop inner diameter 20\% and outer diameter 90\%). The resulting apodizers are displayed in the bottom row of Figure \ref{fig:tel_ap_perim}. In all cases, we find the PSF throughputs of these new designs match those of the circular and keystone APLC to within 2\%. Between these filled hexagonal APLC designs, there is only a small difference in throughput as a function of the number of mirror segments: the Hex-1 aperture has the highest throughput, 23.6\% versus 21.8\% for Hex-4, due to its lower density of segment gaps.

\subsection{Lyot stop geometry investigations}
\label{ls_geometry}

\begin{figure}[!htb]
\centering
\includegraphics[height=8cm]{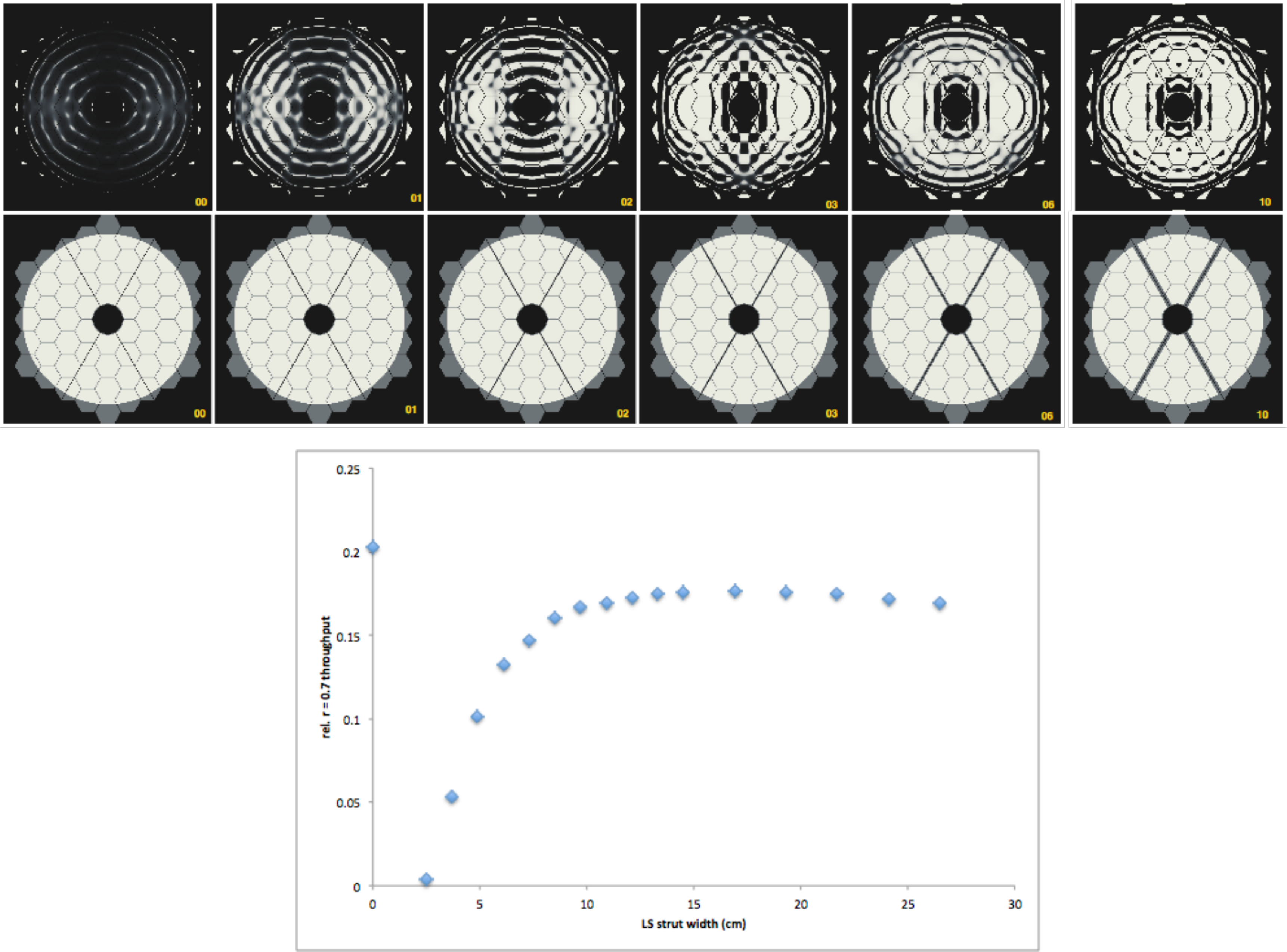}
\caption{Impact of Lyot Stop struts oversize on throughput for the Hex-4 on-axis SCDA aperture. Numbers in yellow indicate the strut oversize in thousands of pupil diameter. The initial strut is 2.5 cm. Throughput plateaus for projected strut size of about 10cm. The first point on the figure is the reference throughput for the APLC (a pupil without spiders)}
\label{fig:LSstruts}
\end{figure}

We have previously established that there is no improvement using reticulated Lyot Stops especially given the added complexity as long as the segment gaps are small enough \cite{2016SPIE.9904E..1YZ}. We also investigated the effect of augmenting the size of the gaps in the numerical aperture used for the APLC optimization. In Figure \ref{fig:LSstruts} we show an important need of slightly over-sizing the Lyot Stop struts. Without over-sizing, the optimizer struggles to find an apodizer solution and the throughput goes to zero. Once the struts are sufficiently oversized the throughput plateaus.

\begin{figure}[!htb]
\centering
\includegraphics[width=12.0cm]{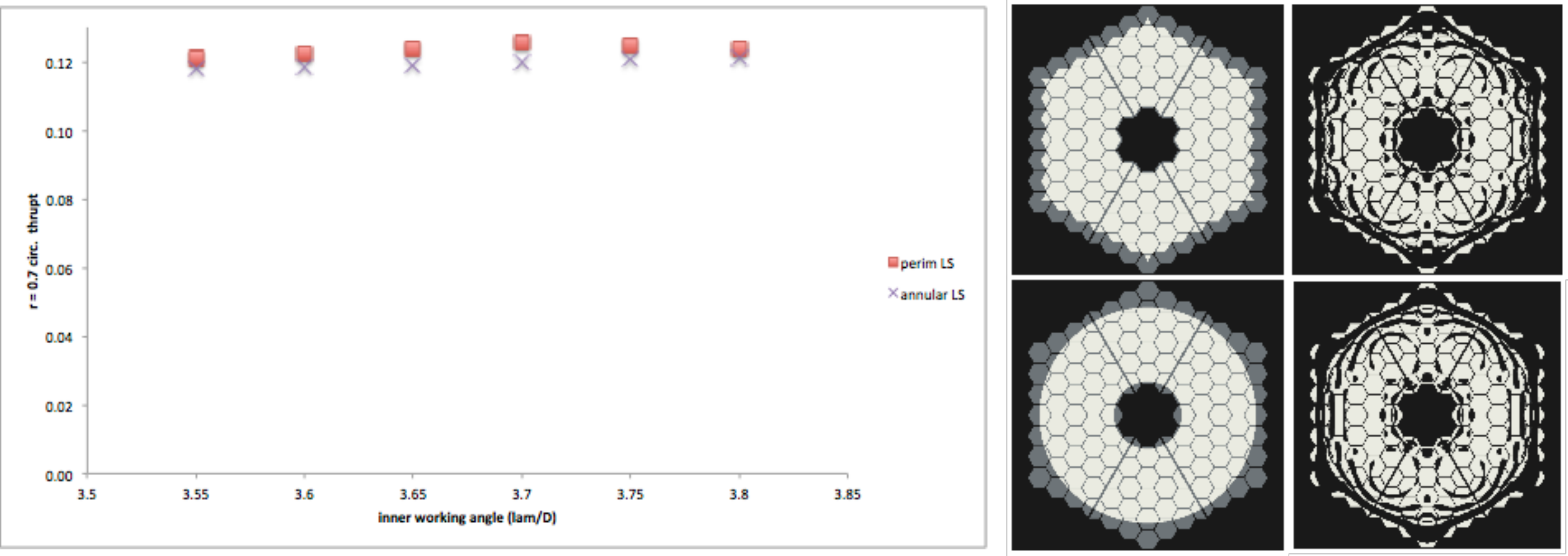}
\caption{Comparison of throughput performance for perimeter and annular Lyot Stop geometries. The figure shows the best combination of ID/OD over the entire survey. There is only a marginal gain for the perimeter LS and no dependency with IWA. The figure shows and example of such two Lyot Stops with their corresponding apodizer, which are very similar.}
\label{fig:perimeter}
\end{figure}

In our previously discussed surveys we use the simplest case of annular Lyot Stop, even when the entrance aperture is not circular. We investigated the case of a Lyot stop geometry that follows perimeter of the hexagonal apertures, which we denote "perimeter Lyot Stop", and applied various amounts of oversizing and undersizing.  For the annular Lyot Stop survey, we considered an inner LS diameter of 20\% to 40\% of the aperture, and an outer diameter from 70\% to 90\% of the aperture diameter. For the Perimeter survey, we used a comparable range of oversize/undersize.  Both surveys explored a range of IWA from $3.55 \lambda/D$ to $3.8 \lambda/D$ and a fixed OWA of $10 \lambda/D$. Figure \ref{fig:perimeter} shows the overall result of this survey, with the conclusion that in the geometry we investigated there is only a marginal gain to consider the perimeter Lyot Stop, and that the simplicity of a circular Lyot Stop is preferable and does not reduce significantly the performance. 

\subsection{Axi-symmetric survey}
\label{axi-sym}

In our first year of study (discussed more fully in \cite{2016SPIE.9904E..1YZ}), we found that APLC performance is independent of the orientations and distribution of thin segment gaps in the primary mirror to first order. Given this, we can extrapolate the behavior of solutions for monolithic circular apertures to segmented apertures, with slight adjustments for decreased apodizer transmission and increased inner working angle. Combining this knowledge with the sharp throughput-IWA transitions discussed in Section \ref{SCDA_ref} and \ref{fpm_rad}, which simplify the selection of FPM radius, we now have an opportunity to identify the jointly optimized FPM size, bandpass, and contrast goal that maximizes the DRM exoEarth detection tallies. This ambitious objective is made practical by the fact that linear programs to design apodizers for axi-symmetric APLCs are orders of magnitude faster than 2-D APLC programs. Therefore, we are poised to answer this question with a relatively small investment of computing time.

To that end we developed axi-symmetric optimization code and integrated it into our scda.py toolkit. In order to fully understand the joint optimization of the aperture, Lyot Stop, and central obscuration we undertook a massive survey of over 65,000 axi-symmetrical designs. In this large and finely sampled survey we varied the central obscuration, lyot stop inner diameter, and lyot stop outer diameter - each defined as a unit-less ratio of telescope aperture. These parameters were varied in steps of 1. Dark zone bandpass was fixed at 10\%  of central wavelenth, dark zone inner working angle at 4.00 \(\lambda/D\), and dark zone outer working angle at 12 \(\lambda/D\).

We found that APLCs can handle a central obscuration up to $\sim25\%$ before the throughput severely degrades. We also found that there is a fair amount of choice available in picking the lyot stop inner diameter, but best performance requires an outer lyot stop diameter at about 93\% for a circular monolith.

\begin{figure}
   \begin{center}
   \begin{tabular}{c}
   \includegraphics[width=14cm]{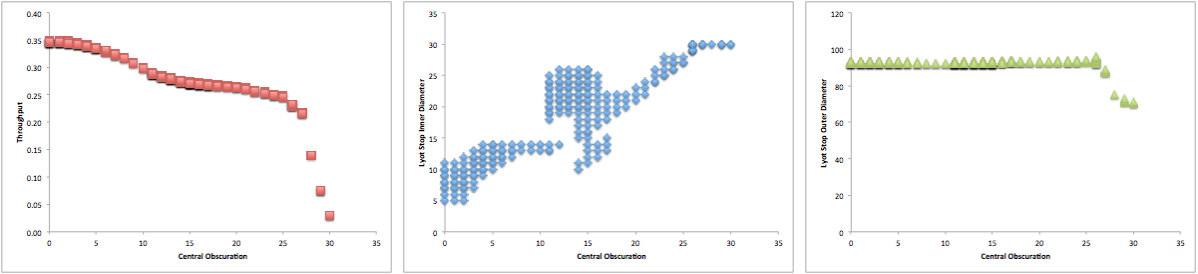}
   \end{tabular}
   \end{center}
   \caption[] 
   { \label{fig:oneD} 
The top 1\% (by throughput) of designs for a given central obscuration were picked out to plot trends. While there is a general decrease in throughput with aperture central obstruction, the APLC can typically handle an aperture central obstruction up to about 25\% before severely degrading the throughput (in the case of a perfect annular aperture). The general behavior is similar with an actual segmented an non-circular aperture, albeit with slightly lower performance. The relationship between inner Lyot stop and central obscuration is somewhat linear, but there are discontinuities for central obscurations between 11 - 14\% of telescope aperture.}
   \end{figure}

\subsection{LUVOIR apertures and parameter space}
\label{LUVOIR_ap_param}

\begin{figure}[!htb]
\centering
\includegraphics[height=7.0cm]{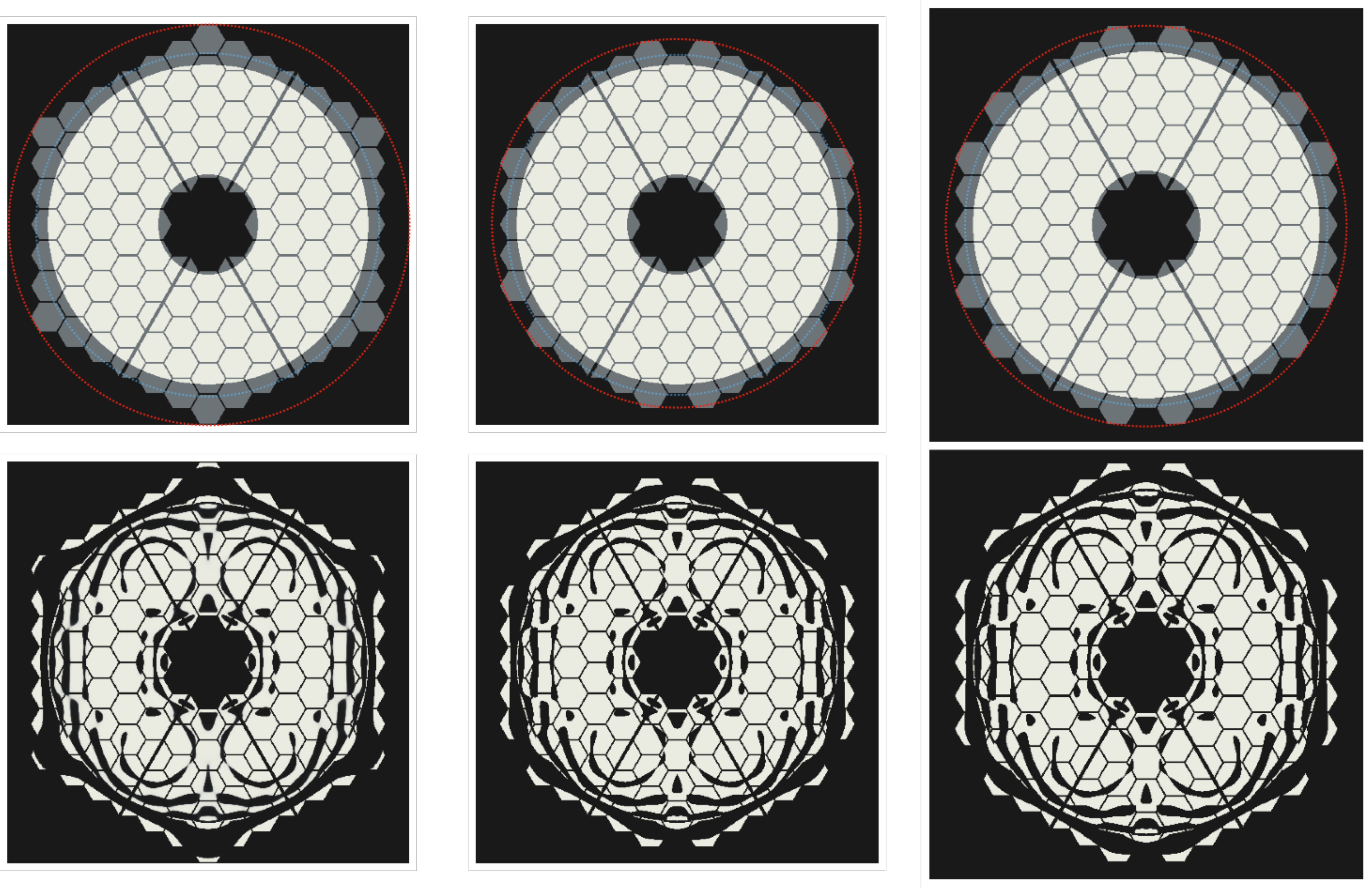}
\caption{Because the APLC apodizer naturally tends to circularize the aperture, corner segments on a hexagonal aperture are mostly black and there is no loss of performance by removing these segments relative to the complete aperture.  Left: LUVOIR Architecture A. Middle: Modified LUVOIR Architecture A with corner segments removed; achieves same performance. Right: Modified LUVOIR Architecture A with same circumscribed diameter as left column (scaling up the segments size); performance is enhanced. This recommendation from our SCDA team has been taken into account by the LUVOIR telescope design team in the new set of apertures.}
\label{fig:cornersegment}
\end{figure}

\begin{figure}[h]
\centering
\includegraphics[height=8.0cm]{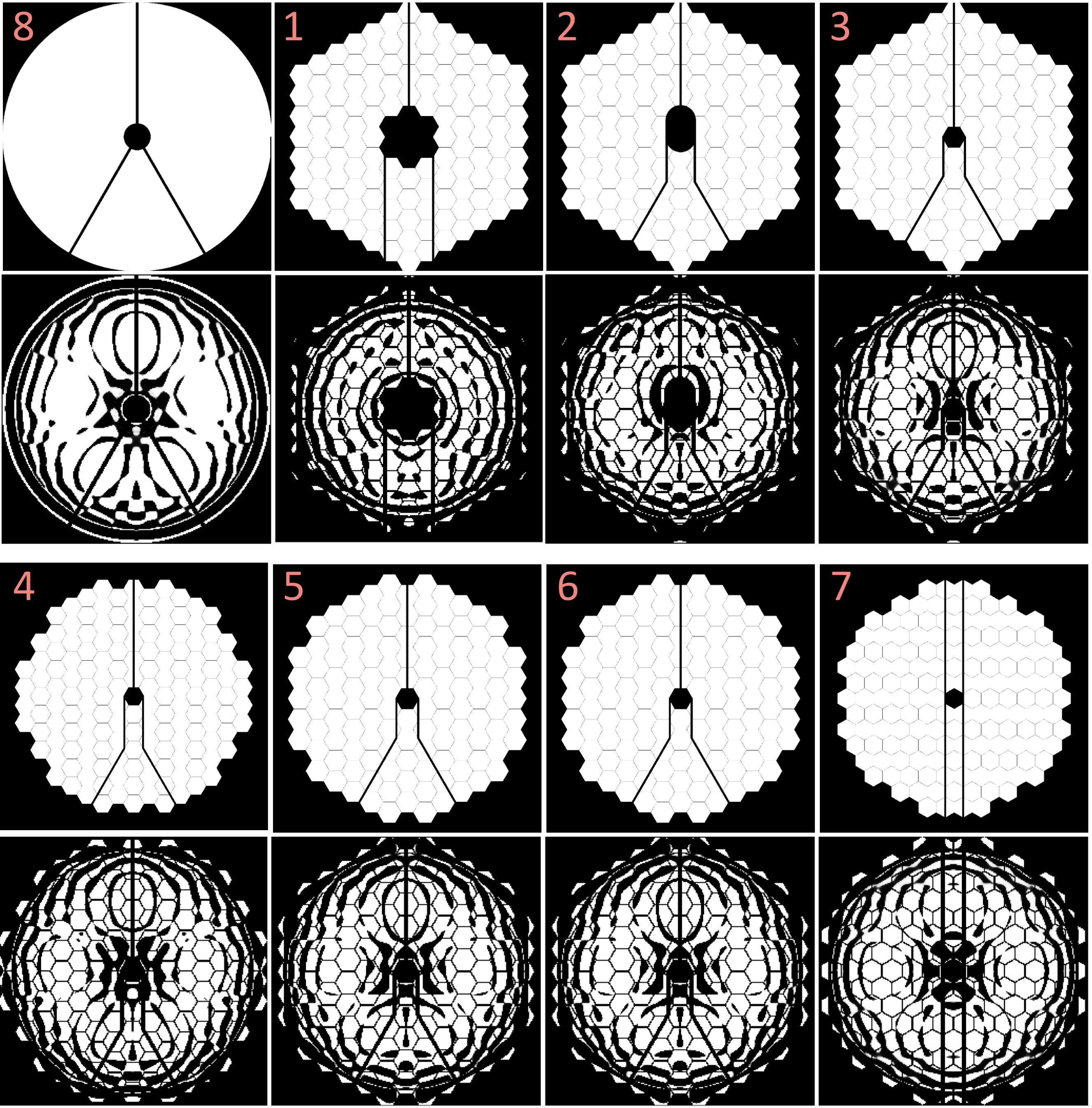}
\caption{Eight telescope apertures surveyed for a LUVOIR "re-design" study, along with their corresponding apodizers for a fixed contrast ($10^{-10}$), dark zone bandpass ($10\%$), and outer working angle ($10\,\lambda/D$) APLC design. On each aperture, the FPM occulter radius (a reliable proxy for inner working angle) was set to its minimum threshold for high throughput, ranging between $3.4\,\lambda/D$ and $3.8\,\lambda/D$. Aperture 1 corresponds to the original ``Architecture A'' pupil, with $15$-meter flat-to-flat diameter. The $14.6$-meter monolithic primary comprising Aperture 8 is used strictly as a performance reference point. Apertures 5 and 6 have the same pupil geometry but are scaled to different physical diameters ($15.2$ m and $16.2$ m respectively).}
\label{fig:redesignapertures}
\end{figure}

As part of our investigation of the LUVOIR "architecture A" aperture design, we have noticed that the corner segments are almost entirely black in the apodizer, and therefore can be removed. Compared to the full aperture this has almost zero loss, but by making the overall aperture diameter smaller, the relative throughput does increase significantly. This is simply because the apodizer naturally tends to circularize the aperture to remove the diffraction features from the aperture geometry.  Therefore, the effective apodized pupil is directly related to the inscribed circle in the telescope pupil. By maximizing the telescope geometry to render it as circular as possible and maximizing the diameter of the inscribed circle with respect to the overall diameter, significant gains can be achieved. One example is illustrated in Figure \ref{fig:cornersegment}. This realization motivated the LUVOIR study office to examine a suite of alternative segmented telescope aperture geometries in order to improve the coronagraph efficiency beyond the original ``Architecture A'' LUVOIR concept. These also included variations in central obscuration size. Thumbnails of these telescope apertures and their corresponding apodizers are shown in Figure~\ref{fig:redesignapertures}. Their relative performances are tabulated in Table~\ref{tab:redesignmetrics}.

\begin{table}[h]
\centering
\small
\begin{tabular}{c||c|c|c|c}
Aperture & Circumscribed & Inscribed & FPM radius   & PSF       \\
         & diam.         & diam.     & (sky angle)  & thrupt.   \\
\hline
$8$ & $1.00$ & $1.00$ & $1.00$ & $1.00$ \\
$1$ & $1.03$ & $0.87$ & $1.09$ & $0.50$ \\
$2$ & $1.03$ & $0.87$ & $1.09$ & $0.62$ \\
$3$ & $1.03$ & $0.87$ & $1.09$ & $0.67$ \\
$4$ & $1.08$ & $0.95$ & $0.95$ & $0.79$ \\
$5$ & $1.04$ & $0.94$ & $1.00$ & $0.77$ \\
$6$ & $1.11$ & $1.00$ & $0.94$ & $0.77$ \\
$7$ & $1.08$ & $0.95$ & $0.95$ & $0.81$ \\
\end{tabular}

\caption{\small Performance metrics of APLC designs optimized for each apertures in the LUVOIR ``re-design'' study (see Figure~\ref{fig:redesignapertures}). All tabulated values are unitless and normalized to the design for Aperture $8$: an obscured monolithic primary of diameter $D = 14.6$ meters, with FPM occulter radius $3.4\,\lambda/D$, optimized for a $10^{-10}$ contrast dark zone with a $10\%$ bandpass and outer working angle $10\lambda/D$, with PSF throughput $22\%$. The throughput is computed as the ratio of the energy inside the unocculted PSF core within radius $0.7\,\lambda/D$ to the total incident energy on the obscured primary mirror.}
\label{tab:redesignmetrics}
\end{table}

\begin{figure}[!htb]
\centering
\includegraphics[width=0.5\textwidth]{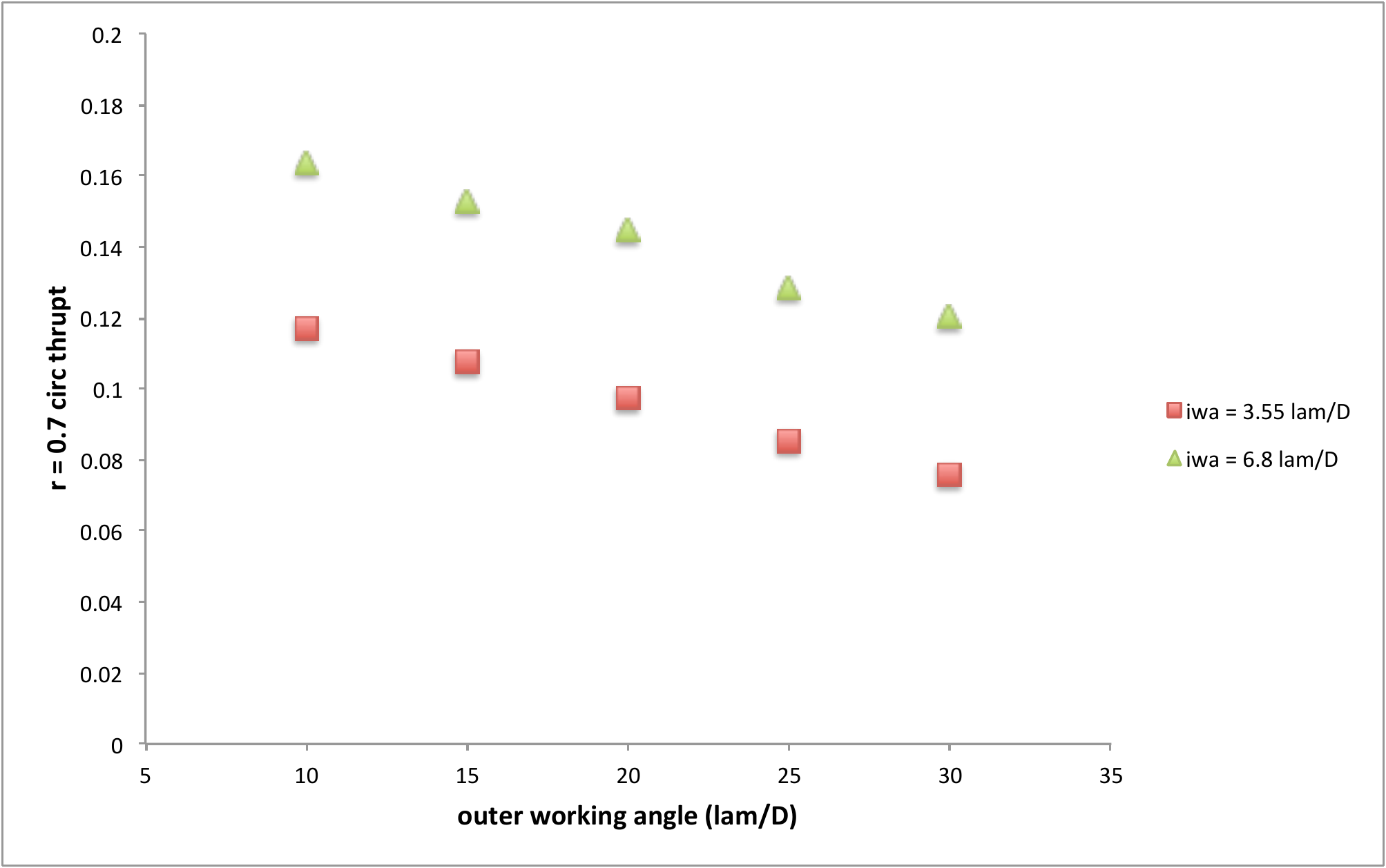}
\caption{Linear dependence of the Airy Throughput with increase OWA. This result seem to be independent of aperture geometry and IWA.  A combination of masks with various IWA/OWA can be derived to optimize the overall science yield of the mission to solve this limitation.}
\label{fig:OWA}
\end{figure}

We investigated the dependence of outer working angle (OWA) vs. throughput. Figure \ref{fig:OWA} shows a linear decrease of the throughput with increase of OWA, assuming a constant IWA.  Thus, a single mask design with small IWA and large OWA, suitable for all stars, would be inefficient.
To solve this, the scientific yield of an APLC instrument would be optimized by choosing a distinct set of masks based on the angular scale of interest for each target star. Since APLC designs with large IWA tend to have higher throughput, a sharper PSF, and enhanced robustness to low-order wavefront error and stellar diameter, a design with relatively high IWA and OWA would be preferable for observing the closest planetary systems, or when observing the outer regions of more distant systems.

We derived a combination of masks that can be paired with any given planetary system, allowing us to observe the full width of any star’s habitable zone while retaining high throughput. Let $H$ be the ratio of habitable zone inner-edge to outer-edge: $H = HZ_{IE} / HZ_{OE} = 0.95 AU / 1.67 AU = 0.56$, and denoting by $X$ the maximum allowable OWA/IWA ratio (based on these results a reasonable range is $X\simeq3-4$), and $IWA_i$ and $OWA_i$ to be the IWA and OWA of the $i^{th}$ mask.  $IWA_0$ should be the minimum IWA that we can design while achieving our contrast goals. The set of IWA and OWAs can then be defined as: 

\begin{equation}
IWA_i = OWA_{(i-1)} \times H
\end{equation}
\begin{equation}
OWA_i = IWA_i \times X
\end{equation}

\section{Robustness of APLC designs}
\label{robust}

\begin{figure}
   \begin{center}
   \begin{tabular}{c}
   \includegraphics[width=12cm]{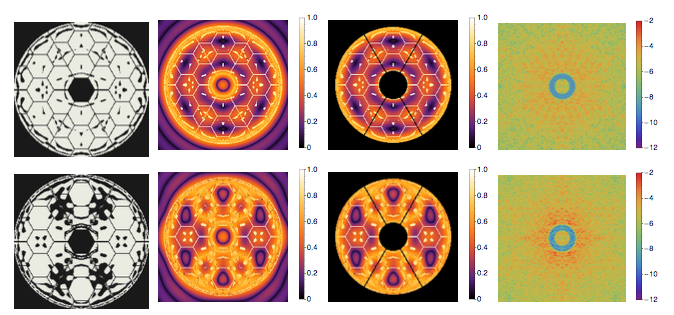}
   \end{tabular}
   \end{center}
   \caption[] 
   { \label{fig:ls_robust_psf} 
   APLC/SP designs with non-robustness (top) and robustness properties (bottom) to Lyot stop misalignments. These coronagraphs are designed for a Hex-3 like aperture with X-like spiders and within an inscribed circle aperture. Optimized for a 4.3 $\lambda/D$  radius FPM, they produce a $10^{-8}$ contrast dark zone ranging between 6 and 10 $\lambda/D$ in the coronagraphic image in monochromatic light. From left: shaped pupil for the APLC, relayed pupil plane image before and after Lyot stop application, and coronagraphic image in the final focal plane showed in log scale.}
   \end{figure}

We approach the development of robust APLC/SP designs by constraining the contrast of the coronagraphic image in a given region for multiple, translated versions of the Lyot stop simultaneously. For our tests, we consider an Hex-3-like geometry aperture with inscribed circle. Our designs are optimized for a 4.3 $\lambda/D$  radius FPM and a Lyot stop with a 2\% undersizing with respect to entrance pupil to produce a $10^{-8}$ contrast in the region ranging between 6 and 10 $\lambda/D$ in monochromatic light. We address two cases: a first design that is solely optimized for a single on-axis Lyot stop and a second design that is optimized for multiple Lyot stops simultaneously (one on-axis and four with offsets positions in x and y axis and in both directions for each axis). The final designs and their light distribution in the different coronagraph planes are displayed in Figure \ref{fig:ls_robust_psf}.

\begin{figure}
   \begin{center}
   \begin{tabular}{c}
   \includegraphics[height=5cm]{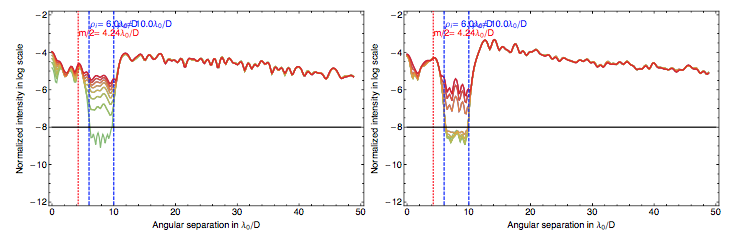}
   \end{tabular}
   \end{center}
   \caption[] 
   { \label{fig:ls_robust_rp} 
   Azimuth averaged intensity profiles of the coronagraphic images with the non robust and robust APLC/SP designs showed in Figure 15, for y-axis Lyot stop misalignments ranging from 0 to 0.89\% of Lyot stop size with a 0.13\% step (from green to red). Intensity is given in log scale. While the non-robust design generates a $10^{-8}$ contrast dark zone only for on-axis Lyot stop, the robust design produces a $10^{-8}$ contrast dark hole for different Lyot stop positions.}
   \end{figure}

Figure \ref{fig:ls_robust_rp} shows the intensity profile of the coronagraphic images for both designs and different Lyot stop positions. On the left plot, the first and non robust design produces a dark hole when the Lyot stop is centered and sees its performance degrading as we move the Lyot stop from its optimal on-axis position. On the right plot, the second design produces dark hole for several Lyot stop offsets, showing ability to yield contrast in the presence of a slightly decentered Lyot stop.

\begin{figure}
   \begin{center}
   \begin{tabular}{c}
   \includegraphics[height=7cm]{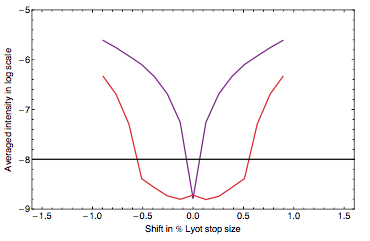}
   \end{tabular}
   \end{center}
   \caption[] 
   { \label{fig:ls_robust_avg_int} 
Averaged intensity of the coronagraphic image in the region ranging between 6 and 10 $\lambda/D$ as a function of y-axis Lyot stop misalignment with the non robust (purple) and robust (red) APLC/SP designs showed in Figure \ref{fig:ls_robust_psf}. Intensity is given in log scale. These coronagraphs produce a $10^-{8}$ dark hole region in the coronagraphic image for Lyot stop offsets of $\pm0.045\%$ and $\pm0.6\%$ of the Lyot stop size. By considering Lyot stop offsets in our design optimization scheme, we improved the tolerance of our APLC/SP coronagraphs by a factor of approx. 10 for a given geometry.}
   \end{figure}

Figure \ref{fig:ls_robust_avg_int} shows the tolerance of both designs to Lyot stop misalignments. Optimized in turn for single and multiple Lyot stop positions, these designs have respective tolerance ranges of $\pm0.045\%$ and $\pm0.6\%$ of the Lyot stop size. In this aperture geometry configuration, the tolerance of our APLC/SP hybrid design is therefore increased by a factor of ∼10 by constraining the PSF contrast for different and simultaneous Lyot stop positions. This result proves very promising in terms of robustness to Lyot stop misalignments but it comes with a cost in apodizer transmission, as we can observe in the shaped pupils in Figure \ref{fig:ls_robust_psf} left frames, and in Airy throughput which drops from 78.2\% to 52.1\%. To recover the throughput while maintaining contrast performance, we explore a second approach, based on the use of wavefront control (WFC) with two deformable mirrors (DMs).

\section{Scientific Yields of APLC}
\label{sci_yields}

C. Stark has previously investigated the dependence of exoEarth detection yield on telescope aperture size and coronagraph performance \cite{2015AAS...22521904S}. More recently he developed a Design Reference Mission (DRM) tool to simulate a space-based coronagraph imaging campaign aiming to detect Earth-like exoplanets. The DRM combines knowledge of nearby star properties, telescope aperture, coronagraph response, and statistical assumptions of the exoEarth planet population, to formulate a survey plan (target star and visit sequence) optimized for the provided telescope and coronagraph combination. The result is an estimated exoEarth yield after 1 total year of exposure time (1 additional year budgeted for overheads). In our present simulation, the yield is for V-band detections only. The DRM did not devote any time to spectral characterization, since we would use a different mask design for characterization. Because yield is a weak function of total exposure time, we don’t expect characterization time to dramatically change the relative performance of the various designs.

\begin{figure}
   \begin{center}
   \begin{tabular}{c}
   \includegraphics[width=12cm]{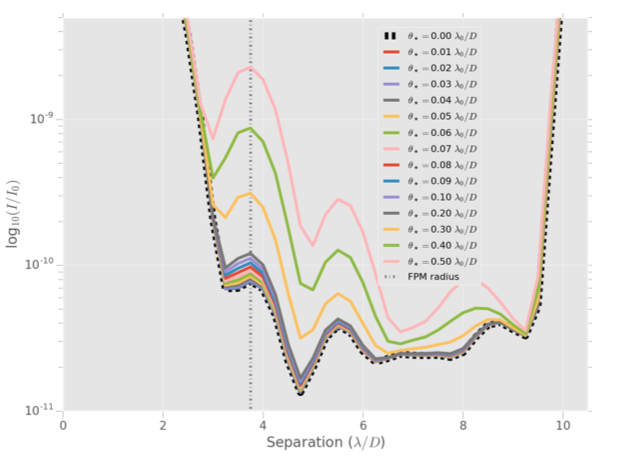}
   \end{tabular}
   \end{center}
   \caption[] 
   { \label{fig:SCDA_ref_sy} 
The nominal radial profile of the on-axis PSF (averaged over azimuth) computed for a range of stellar angular diameters. This is an APLC design for centrally obscured Hex-4 aperture with FPM radius fixed at 3.75 $\lambda_{0}{0}/D$. For reference, at wavelength  $\lambda_{0} = 600 nm$ for a 12-meter diameter primary mirror, star diameter $\theta_{*} = 0.1$ corresponds to approximately 1 milliarcsec, or the angular diameter of a solar twin from a distance of 10 pc.}
   \end{figure}

We evaluated the estimated exoEarth yield of the solutions produced by our design survey, taking the highest throughput design at a given IWA and bandwidth. As input to the DRM, we specify the coronagraph response using a standardized suite of two-dimensional PSF modeling products. These products take into account the gradual degradation of the dark zone contrast with the increasing star diameter (illustrated in Figure \ref{fig:SCDA_ref_sy}). We did not include wavefront aberrations or wavefront control, as such features are not yet incorporated into our toolkit. Therefore, the PSFs were determined strictly by Fourier propagation of plane waves through the coronagraph mask train.

Table \ref{table:scda_ref_sy} lists the number of detected exoEarths in V-band for the best APLC design for each combination of telescope aperture type and diameter. For the obscured/on-axis telescope architectures, the designs with peak yield are easily identified in the throughput-IWA profiles in Figure \ref{fig:throughput_SCDA}, corresponding to the point at the cusp of the throughput “plateau”: FPM radius 3.75 $\lambda_{0}/D$ for the hexagonal apertures and 3.25 $\lambda_{0}/D$ for the keystone/pie-wedge apertures. For the unobscured/off-axis telescope architectures, the designs with the highest number of exoEarth detections have FPM radius 3.50 $\lambda_{0}/D$ for the hexagonal apertures and 3.25 $\lambda_{0}/D$ for the keystone aperture.

For all of the apertures, the best 10\% bandpass design outperformed the best 15\% bandpass design. Therefore, all the yield values in Table 1 correspond 10\% bandpass designs. This result suggests that the expanded discovery space owed to the reduced inner working angle tends to outweigh the loss in band-integrated count rate. It also underscores the fact that the APLC bandpass that maximizes the number of rocky planet detections could be significantly narrower than the bandpass limit imposed by the wavefront control system performance. In the future, we will investigate how APLC yield varies for a wider assortment of bandpass values, testing bandpasses narrower than 10\% as well as intermediate values between the 10\% and 15\% bandpass cases we covered in 2016.

Some prominent nearby stars evaluated by the DRM simulation, such as $\alpha$ Centauri, have such large apparent angular diameters (i.e., 7 milliarcsec) that the program excludes them from the optimized survey plan, and they make no contribution to the detection tally. However, such targets, which are of particularly high scientific value due to their proximity and potential for extensive characterization, are not necessarily inaccessible to the apodized/shaped pupil coronagraph design family. If we allow for an instrument architecture that can switch between multiple coronagraph masks, then we could design a specific mask that performs advantageously on nearby, high angular diameter stars. Because the inner working angle required to search for habitable exoplanets around the most nearby targets is relaxed, the solution could be as simple as a conventional, 2-D optimized shaped pupil without a Lyot mask train \cite{2011OExpr..1926796C}.

One striking feature of Table \ref{table:scda_ref_sy} is that the detection tallies for the obscured/on-axis telescope architectures are generally within a factor of 2 or less of the unobscured/off-axis yields. For the smallest simulated mission apertures (4.0 meters and 6.5 meters), the tallies for all aperture cases converge to within 4 detected exoEarths. At the large aperture end, where the impact of the central obscuration is highest in terms of absolute detection tally, in terms of detection proportion the disparity is at most 24\%, and in one case as low as 4\%.

Also of interest, Table \ref{table:scda_ref_sy} shows that the keystone APLC performs very close to the reference ``ceiling'' set by the circular monolith APLC, and in fact exceeds the circular monolith APLC DRM detection tally by 1 planet for the 16-meter aperture with central obscuration. This reversal of performance rank  contradicts what we would expect on the basis of throughput metrics alone. However, it can be explained by the pseudo-random variation in the responses of individual APLC designs to stellar diameter. Stellar diameter tolerance is controlled only indirectly by our apodizer optimization procedure. Therefore, the PSF degradation (shown for one example in Figure \ref{fig:SCDA_ref_sy}) is numerically sensitive to the specific input parameters of the linear optimization program, and for the time being we have no way to precisely predict its outcome. Therefore, two APLC designs with similar telescope apertures, apodizers, and performance parameters can produce DRM simulations with marginally different detection tallies.

\begin{table}[]
\centering
\begin{tabular}{cc||c|c|c|c|c}
\multicolumn{1}{l}{}            &            & \multicolumn{5}{c}{Diameter (m)} \\
\multicolumn{2}{c||}{Aperture type}           & 4.0  & 6.5  & 8.0  & 12.0  & 16.0 \\ \hline \hline
\multirow{2}{*}{Hex 1}          & Obscured   & 3    & 8    & 11   & 22    & 37   \\
                                & Unobscured & 5    & 11   & 16   & 31    & 47   \\ \hline
\multirow{2}{*}{Hex 4}          & Obscured   & 3    & 9    & 13   & 26    & 44   \\
                                & Unobscured & 4    & 10   & 14   & 28    & 46   \\ \hline
\multirow{2}{*}{Keystone 24}    & Obscured   & 4    & 11   & 15   & 31    & 52   \\
                                & Unobscured & 5    & 12   & 18   & 36    & 59   \\ \hline
\multirow{2}{*}{Circ. Monolith} & Obscured   & 4    & 10   & 15   & 31    & 51   \\
                                & Unobscured & 5    & 13   & 19   & 39    & 63  
\end{tabular}
\caption{ExoEarth detection yield metrics of the best SCDA APLC solutions for each telescope aperture (varying segmentation pattern and inclusion of central obscuration). The yields are determined by C. Stark’s DRM code, modeling detections in V band only.}
\label{table:scda_ref_sy}
\end{table}

Worth emphasizing absolute numbers are subject to change (instrument details, eta-Earth, other assumptions), but relative patterns should remain consistent.

We continue to evaluate the exoplanet science yield of APLC masks from the SCDA study using the yield input standards that resulted from this work (Stark \& Krist, \href{www.starkspace.com/yield_standards.pdf}{www.starkspace.com}).  

Recent improvements to the code, funded externally to the SCDA study, have included an updated input target list (with improved stellar fluxes, increased completeness, and better binary parameters), a more detailed treatment of stray light due to binary companions, an improved detector noise treatment, updated planet occurrence rates, the ability to handle multiple types of planets, and the ability to handle multiple coronagraphic masks by pairing a given coronagraph with a given star.  This last feature has been crucial to estimating the relative performance of different telescope/coronagraph designs given the new multi-mask approach described in Section 4. We find that adding larger IWA/OWA masks for closer/earlier type stars can increase yield by 25\% or more.

We have also used the yield optimization code to evaluate the performance of APLC masks designed for several LUVOIR Architecture A options.  This tool has allowed us to ``close the loop," by tracing changes in the aperture geometry to impacts on coronagraph design, and to scientific yield. The yield code has guided the redesign of the LUVOIR Architecture A and has showed that the lessons learned in the SCDA study (described above) have a substantial impact on LUVOIR’s yield: decreasing the size of the secondary while circularizing the outer perimeter of the mirror by removing corner segments can roughly double the exoplanet science yield.

\pagebreak
\section{Conclusions}

APLC performance is mainly constrained by the presence and size of the central obscuration and the deviation of the primary mirror perimeter from a circle. The struts and segment gaps considered by SCDA are geometrically thin enough (generally $\sim 1\%$ or less of pupil diameter) such that performance is only weakly affected by the specific segmentation pattern within the telescope pupil.

The impact of the central obscuration on throughput and inner working angle is significant but does not pose a fundamental threat to mission objectives. In our preliminary design reference mission (DRM) yield analysis, the number of exoEarths detected with an obscured 12-meter diameter telescope is within 30\% of the number detected by the same telescope without central obscuration. However, the central obscuration considered in our study is relatively small (10\% of pupil diameter). We expect the performance to deteriorate quickly with increased obscuration fraction, a difficulty the WFIRST coronagraph designs contend with, having an obscuration 31\% of pupil diameter.

We identified sharp transitions in APLC throughput as a function of inner working angle,in particular for centrally obscured apertures. These transition points shift depending on the dark zone bandpass (as indicated in Figure \ref{fig:throughput_SCDA}), so that designs for a larger bandpass will generally demand a larger FPM. There is a clear advantage to operating with an FPM matched to the inner edge of this throughput-IWA plateau; these are the designs which maximize the exoEarth detection yields listed in Table \ref{table:scda_ref_sy}.


\acknowledgments 

This work was supported by Jet Propulsion Laboratory subcontract No.1539872, funded by NASA and administered by the California Institute of Technology. (PI R. Soummer). .Resources supporting this work were provided by the NASA High-End Computing (HEC) Program through the NASA Center for Climate Simulation (NCCS) at Goddard Space Flight Center (awards SMD-16-6657, SMD-16-6962, and SMD-16-7146). We are grateful for the technical support we have received from Nick Acks and other NCCS staff members.

\bibliography{bibliotheque}
\bibliographystyle{spiebib}

\end{document}